\documentclass[preprint]{aastex}
\begin{document}
\newcommand{\etal}{{et~al.\/}\ }
\newcommand{\eg}{{\em i.e.\/}\ }
\newcommand{\ie}{{\em i.e.\/}\ }
\newcommand{\cf}{{\em c.f.\/}\ }
\newcommand{\etc}{{\em etc.\/}\ }
%
\newcommand{\sgs}{\sigma^2}
\newcommand{\ham}{H{\alpha}}
\newcommand{\ha}{$\rm H{\alpha} $ }
\newcommand{\mh}{ \cal  M_{\rm HI}  }
\newcommand{\mo}{ \cal  M_{\odot}  }
\newcommand{\kms}{$\rm km \ s^{-1}$}
\newcommand{\hl}{\hspace*{1in}}
\newcommand{\hlp}{\hspace*{1.1in}}
\newcommand{\hs}{\hspace*{.5in}}
\newcommand{\vs}{\vspace*{3in}}
\newcommand{\nvs}{\vspace*{-2mm}}
\newcommand{\ivs}{\vspace*{-3mm}}
\newcommand{\sml}{\smallskip \\}
\newcommand{\fil}{ {\em fill} \hspace{.5in} }
\newcommand{\ph}{$\rm \phi$}
\newcommand{\ho}{$ \rm h^{-1} $ }
\newcommand{\g}{GC's }
\newcommand{\s}{GCS }
\newcommand{\sn}{\(\rm S_N \) }
\newcommand{\slu}{\(\rm S_L \) }
\slugcomment{Paper II: Submitted to  A.J.}

\shorttitle{NGC~3256: H~I}
\shortauthors{English et al.}

\title{NGC~3256: Kinematic anatomy of a merger}

\author{J. English}
\affil{University of Manitoba}
\authoraddr{Department of Physics and Astronomy, University of
Manitoba, Winnipeg, Manitoba, Canada R3T 2M8}

\author{R. P. Norris}
\affil{Australia Telescope National Facility}
\authoraddr{CSIRO Radiophysics
         Laboratory, P.O. Box ~76, Epping, NSW , Australia} 

\author{K. C. Freeman}
\affil{Mount Stromlo and Siding Spring Observatories}
\authoraddr{Private Bag, Weston Creek P.O., 
A. C. T. ~2611, Australia}

\and

\author{R. S. Booth}
\affil{Onsala Space Observatory} 
\authoraddr{Chalmers University of Technology, S-439 ~00, 
         Onsala, Sweden }

\begin{abstract}

We have used the Australia Telescope Compact Array to image
 the neutral hydrogen in the merging system
NGC~3256, to test
 the idea that globular clusters (\g)
 form during the interactions
and mergers of disk galaxies. 
We compare our observations with hydrodynamical numerical simulations,
from the literature, to examine the hypothesis
that  the H~I fragments with masses greater 
than 10$^{7\pm1} \mo$ are sites of GC formation.  
We appear to have
detected  detached H~I fragments in the vicinity of NGC~3256. 
These fragments, with masses $\rm \sim 10^7 \mo$, may have 
little dark matter content which is also a characteristic
of globular clusters, and so our observations support the hypothesis
that globular clusters form in the type of interaction that 
resulted in NGC 3256.

\end{abstract}


\section{Introduction}
\label{intro}

NGC~3256 is a starburst galaxy in which 2 extended tidal tails reveal
that an interaction between galaxies has occurred (see
Fig.~\ref{postcard} and~\ref{4bd}). It is included in the description
by \nocite{toom77} Toomre~(1977) of the sequence of stages associated
with merger of 2 disk galaxies.  Kinematic information about the gas
in such a system could help answer questions such as whether mergers
trigger starbursts, whether starbursts play a powerful role in active
galactic nuclei, and whether mergers create gas condensations which
could evolve into globular clusters (\g). Since NGC~3256 is not in a
cooling flow group or cluster, and cooling flows are also suspected of
playing a role in the formation of globular clusters, it is a key
galaxy for studies which are tackling these questions.

Galaxy-galaxy interactions, which are a mechanism for driving
reservoirs of hydrogen gas from the outer regions of the galaxies
into the centre of the interacting system, trigger starbursts.
This picture is consistent with the
observation that CO emission, which indicates the presence of
$\rm H_2$, is strong and centrally concentrated in perturbed
galaxies. Additionally galaxy-galaxy interactions could provide 
high pressure environments conducive to GC formation.   For example,
tidally shocked debris torn from the parent galaxies could have
characteristics similar to the giant clouds which are thought to be 
proto-galactic building blocks.   Bound clusters, such as GC, 
preferentially form in the dense, high pressure cores of
such massive clouds (eg. \citet{harpud94}, \citet{elm97})

In the remainder of this section we will briefly discuss some
aspects of nuclear activity and globular cluster formation in the
context of the merging of two disk galaxies 
to form an elliptical-like galaxy.  We also describe the
known characteristics of the merging system NGC~3256.

Observations are presented in \S\ref{obs}, \S\ref{results}
presents the results and \S\ref{gcres} analyses the H~I
enhancements suspected of being detached gas fragments,
\S\ref{tail} compares the global features evident in our data
(described in \S\ref{obs}) with some numerical models of galaxy
mergers from the literature, and   \S\ref{discus} reviews the
characteristics which {\em suggest} that the detached fragments
could be GC progenitors.

\subsection {Globular Cluster Formation} 
\label{gcform}
It is now widely believed that the dominant giant ellipticals in
clusters of galaxies are produced by mergers during the dynamical
evolution of the cluster.  The number of globular clusters per unit
luminosity (the specific frequency \sn) for  these cluster giants is
higher, by up to an order of magnitude, than for other ellipticals and
spirals \citep{har93}.  Based on observations of starburst activity
and the large amounts of molecular gas in interacting galaxies,
\nocite{sch88} Schweizer~(1988) argues that merger events provide an
ideal environment for the formation of massive star clusters. Hence the
\sn  of an elliptical could be larger than the value derived from the
sum of the \g populations of the parent galaxies.  The specific
frequency issue is discussed in more detail in English \& Freeman
(2000; Paper I).

In Paper I, we focused on a scenario based on \nocite{kc88} Kennicutt
and Chu's (1988) demonstration that giant H~II regions are probably the
birthplaces of the young globular clusters (young populous clusters;
YPC) as seen in the Magellanic Clouds. Our argument is that tidal
disturbances provide a mechanism for enhancing star formation and hence
generate giant H~II regions in the disks  of the parent galaxies and in
the newly emerging galaxy. These ionised complexes in turn cradle YPC's
which are the young GC's.  The new \g are redistributed, along with the
other components of the two disks involved, during subsequent changes
in the potential field and hence the clusters in  the elliptical have a
roughly similar spatial distribution to the diffuse light distribution
(as observed  by \nocite{hh87} Harris and Hanes~1987, for example).

Here we examine an extension of \nocite{sz78} Searle \& Zinn's~(1978)
proposal that globular cluster formation occurs in fragments around
collapsing protogalaxies:  gas fragments and also condensations in the
tails and loops of merging systems  experience tidally induced shocks,
causing the formation of stellar subsystems.  These fragments, together
with the globular clusters that formed within them, subsequently become
part of the final merged galaxy system. (We elaborate on this scenario
in \S\ref{discus}.) In this paper, we describe a search for such HI
fragments around the merging system NGC 3256.

\subsection {NGC~3256}
\label{intro3256}
NGC~3256 consists of two galaxies which are currently merging.  It is
very luminous in the far infra-red ($\rm L(8-1000\mu m)\sim 3 \times
10^{11}\ L_{\odot}$~\citep{ssp89}; $\rm L(10 \mu m) \sim 3 \times
10^{10}\ L_{\odot}$~\citep{grah87}), although not quite in the class
of FIR ultra-luminous  galaxies ($\rm \geq 10^{12}\  L_{\odot}$).

Its starburst nature is indicated by data over a wide wavelength range:

\begin{itemize}
\item The UV spectrum (\nocite{kin93} Kinney et al.\/~(1993)) has
strong absorption features which indicate the presence of hot young
stars;

\item The [Fe II] and radio continuum data (\nocite{nf95} Norris \&
Forbes~(1995)) indicate that the radio emission is due to synchrotron
emission from cosmic-ray electrons accelerated by supernovae whose
progenitors were massive stars.

\item Infrared observations by \nocite{grah84} Graham et
al.\/~(1984), \nocite{djw94} Doyon, Joseph, \& ~Wright~(1994), and
\nocite{mo94} Moorwood \& Oliva~(1994) indicate that the K band
continuum emission and CO strength are due to red supergiants and
that the high $\rm Br_{\gamma}$ line emission is due to OB stars
ionising H II regions.

\item \nocite{sk93} Smith \& Kassim~(1993) show that the spectrum
of NGC~3256 closely resembles the archetypal starburst galaxy M~82 
over the wavelength range from about 30 cm  to  7500
\AA.  \end{itemize}

We also note that although NGC~3256 has relatively few (7) giant
H~II regions, these regions are comparable in flux to about 36
30-Doradus-like H~II regions (Paper I).  Hence this system
currently has a giant H~II region formation rate 30 times that
observed for typical Sc galaxies with $\rm M_B = -19.5$~\citep{kc88}.
Additionally \nocite{moran99} Moran et al. (1999) find that it
is the most X-ray-luminous star-forming galaxy yet detected ($\rm
L_{0.5-10keV} \ = \ 1.6 \times 10^{42}\ erg \ s^{-1}$).

As discussed in \S\ref{tail}, the appearance of the stellar tidal tails
of NGC~3256 implies that the initial interaction event involved two 
similarly massive disk galaxies,  and the luminosity of the stellar
envelope of this merging system  implies that the end product will be
an elliptical galaxy. At this stage of the merger process, the system
appears to have two nuclei. One nucleus (northern) is visible
in the optical continuum and H$\alpha$.  The other nucleus,
obscured in optical images, is present in the [Fe II] and
radio continuum (3 and 6 cm) images of \nocite{nf95}Norris \&
Forbes~(1995).  It is also evident in the K band images of
\nocite{zl93} Zenner \& Lenzen~(1993), \nocite{mo94}Moorwood \&
Oliva~(1994), and \nocite{mcg94} McGregor \&  Kim~(1994). The upper
image in Fig.~\ref{4bd} is composed of our \ha continuum-subtracted
data (see Paper I), \nocite{nf95}Norris \& Forbes 3 cm data, and
McGregor \&  Kim's  K-band data, which is also displayed in the lower
image.

The identification of the two strong emission sources as the two
nuclei of the parent galaxies is supported by the following
arguments:
\begin{itemize}
\item In the radio continuum data,  the fluxes 
and the spectral indices
of these features are comparable~\citep{nf95}.
\item \nocite{mk94} McGregor \& Kim~(1994) 
have found that the southern feature is 1.5  
magnitudes fainter
than the \ha nucleus in K, and its J-H and H-K colours  are redder.  
This is consistent with 
the notion that the southern feature is a galaxy 
nucleus suffering foreground extinction.
\item Using Chandra X-ray Observatory \citet{lira02} identify
an X-ray object with each source, and using resolved NICMOS IR images,
argue that the southern source is a nucleus with about 16 magnitudes
of extinction.
\item Although \nocite{grah87}Graham et al.\/~(1987)
claim that the K-band surface brightness distribution can be
described by an $\rm r^{1/4}$ profile out to a radius of 5 kpc,  the 
surface brightness profile from the  K image of 
\nocite{mo94}Moorwood \& Oliva~(1994) 
is not that of a galaxy which has already relaxed sufficiently
to be classified as an elliptical. 
\item \nocite{ssp89}Sargent et al.\/~(1989) determined that the
far infra-red luminosity per unit $\rm H_2$ mass is low
relative to the FIR ultra-luminous galaxies.  
They argue that this value, together with the large extent of the CO
emission and the unusually high mass of molecular gas, indicate
a relatively early merger stage, in which the galaxy cores have not
yet coalesced.
\end{itemize} 
 
The near-infrared data  also demonstrate that the 
starburst activity is extended.    
The  two nuclei, the eastern \ha
complex,  and features
in the northern \ha ``arc''  and in the southwest projection all have
$\rm Br_{\gamma}$ emission.  The photometry of  \nocite{grah87}
Graham et al.\/~(1987) indicates that most of the IR luminosity
arises outside the central kpc (i.e. the \ha nucleus)
of the system. 

In Paper I we examined that possibility that giant H~II regions could
be sites of GC formation. Our analysis suggests that the number of
YPC's produced is  consistent with  specific frequencies of \g
estimated for elliptical merger remnants  formed via mergers of spiral
galaxies~\citep{az93}.  In the remainder of this paper we examine
whether fragments of neutral hydrogen exist around NGC~3256 which could
also be future sites of GC formation, without presuming a link between
this scenario and the one in which \g are cradled in H~II regions
(\S~\ref{gcform}).

\section{Observations, Reductions, and Analysis}
\label{obs}
\subsection{Single-Dish Observations}
A single-dish HI profile (Fig.~\ref{pksprof})
was obtained using the ATNF's Parkes
Radio Telescope
 on March 2,~1991;
at 21-cm the FWHM (full-width; half-maximum) of the beam 
is 15 arcmin. The system temperature was
calibrated using Hydra A (43.5 Jy).
We obtained two spectra with orthogonal 
polarisations, each with 512 frequency channels over 10 MHz bandwidth, 
for a total on-source integration time of 30 minutes  and a reference 
integration time of 30 minutes.
After fitting baselines to individual
input scans, summing these to produce the spectra,
and applying Hanning smoothing, 
the r.m.s. was 6 mJy
and the velocity resolution was about 8 \kms.
 
\subsection{Radio Synthesis Observations}
Table~\ref{atcapar}
lists parameters associated with the spectral-line 
data cube described below.

The 21-cm HI synthesis data was acquired
using three
different antenna configurations of   the
Australia Telescope Compact Array (ATCA). 
The shortest baseline corresponded to an angular
resolution of about 23 arcmin while the longest was about
7 arcsec with the choice of configurations 
optimized for angular scales larger than 30 arcsec. We
observed the source on December 17, 1991, January 10,
1992, and June 18, 1992 for about 12, 10 and 7 hours
respectively, with a secondary  calibrator (PKS1104-445) observed for 5
minutes every 30 minutes. We used 512 channels over a
bandwidth of 8 MHz 
centred on a heliocentric
velocity of 2770 \kms, although these were later averaged
in groups of five channels to produce a final velocity
resolution of 16.8 \kms. 

We processed the data using the Australia Telescope
National Facility version of the AIPS reduction package.
Each configuration was separately flux calibrated,
adopting a flux density of 16.21 Jy 
(at 1407 MHz) for the primary calibrator
source 1934-638. The continuum was subtracted in the U-V
plane with UVLSF, which is a routine that fits to the real and
imaginary parts of visibilities associated with channels
designated as containing continuum-only emission. The
data from the three observations were then shifted to a
common velocity 
 before they were combined to produce the UV data set.  
A subset of this set was 
mapped and cleaned using the AIPS
task MX, to produce a spectral line cube of 33 image planes
(each plane being a group of 5 channels) covering a velocity
range of 555 \kms . 
Each plane has a typical  
rms noise of 0.9 mJy/beam (after primary beam correction),
spans 512 $\times$ 512 pixels, and the mean diameter
of the synthesised beam 
23 arcsec.

The background in each plane contains 
imaging artifacts consisting of randomly distributed
rings that have the same scale  as the 
innermost grating ring of the synthesised dirty beam. 
We did not self-calibrate the data since this technique
 is unreliable for ATCA observations on sources as weak 
and extended as NGC~3256. 
 
A condensed version of this cube was made for 
display purposes; see Fig.~\ref{mosaic} and \S\ref{velbehav}.  
The cube was also smoothed for comparison with the Parkes
single-dish data; see \S\ref{emis}. 
\begin{table}
\dummytable\label{atcapar}
\end{table}

\section{Results}
\label{results}

\subsection{Global properties of NGC~3256}

The integrated surface brightness map is shown Fig.~\ref{HIcntr}; see
the figure caption for details.  
While Fig.~\ref{HIcntr}a was constructed
after correcting for the primary beam, Fig.~\ref{HIcntr}b was
generated beforehand and indicates whether any features could be
artifacts generated by this correction.

Fig.~\ref{postcard}, which shows the relationship between H~I
emission (coloured red) and optical I-band emission (from
Paper~I), was produced at a time when only two configurations of
the antenna array were available. The I band image (coloured
green) is representative of the older stellar population and the
continuum-subtracted ionised hydrogen image (coloured blue)
displays star forming regions.  This figure emphasizes  a
neutral hydrogen absorption feature which almost covers the
spatial extent of the optical disk.

\subsubsection{21-cm  Flux Density}
\label{emis}
The single-dish spectrum is presented in Fig.~\ref{pksprof}; the integrated 
flux density from this spectrum is 18 $\pm$ 1 Jy \kms. Since this is an 
underestimate of the HI emission, because it includes the central absorption 
feature, we measured the emission-only flux density from the spatially
resolved ATCA data. Using the data in Fig.~\ref{HIcntr}, we integrated 
the emission in annuli that were centred on the absorption feature of
NGC~3256 and which had an annular width equal to the size of the
major axis of the Parkes beam, and excluded the inner 26 arcsec of the
galaxy from these measurements. We did not correct
for inclination.  The resulting integrated flux density of 21-cm
emission is 19 Jy km s$^{-1}$, with an estimated 10\%
uncertainty.

As a further check, we compared the Parkes single-dish spectrum with
a spectrum obtained by convolving the
ATCA data  with a Gaussian equivalent to the Parkes single dish beam.
The two spectra are shown
in Fig.~\ref{emprof}. 

From the ATCA data, we also measured the emission in each of the
tails, excluding apparent fragments. To avoid the large velocity width
features evident in the velocity curve (Fig.~\ref{HIrc}), we assumed
the base of each tail resided beyond the 70 arcsec radius associated
with the I band stellar envelope (see Paper~I).  This approach
excluded emission which may belong to the east tail (i.e. the H~I
residing south of the absorption feature) and hence the value of 4.6
Jy km s$^{-1}$ could be underestimated up to 30\%. The value for the
west tail, 9.6 Jy km s$^{-1}$, is larger but may also be an
underestimate. Therefore at least 75\% of the H~I emission comes from
these extended features.

\subsubsection {Velocity Field}
\label{velbehav}

Due to the peculiar morphology of NGC~3256, and the extinction in the
centre, it is difficult to determine the centre of this system, its
major axis and its inclination to the line of sight.
\nocite{fr78}Feast \& Robertson (1978) used \ha spectroscopy to
determine a position angle of $100^{\circ}$ for the major axis.  They
also adopt a kinematic minor axis which is offset from the visible
nucleus by 4.5 arcsec to the west, with an associated systemic
velocity of 2820 \kms .  Our \ha long slit spectrum, described in
Paper~I, with the slit at position angle $90^{\circ}$, also gives a
systemic velocity of 2820 ($\pm$ 11 \kms) and a rotational velocity
amplitude of about 107 $\pm$ 11 \kms, uncorrected for inclination.
This systemic velocity is associated with the disk which appears less
disturbed by the merging activity than do the H~I arms of the galaxy
and the amplitude is significantly larger than the separation of the
H~I horns in the ATCA H~I emission profile, Fig.~\ref{emprof} (which
gives an H~I systemic velocity estimate of 2813 \kms).

Fig.~\ref{mosaic} is a mosaic of contour plots of the HI
emission, which has been corrected for primary beam effects. Each
panel consists of two planes of the cube that have been summed
together with a resultant velocity width per panel of about 34
\kms \ and a noise level of 1.3 mJy. We discuss the evident
non-circular motions in \S~\ref{gcres}.

A map of intensity weighted velocity is
shown in Fig.~\ref{HIvfld}, and in Fig.~\ref{HIrc} we present
an H~I position-velocity diagram at the same
declination and position angle as the \ha long slit spectrum. The
velocity behaviour in H~I shows overall rotation in the same
sense as that described by the optical velocity map presented in
Paper~I.
 
\subsubsection {HI Absorption}
The neutral hydrogen gas distribution in the centre of the galaxy
is characterised 
by a strong spatially-unresolved
 absorption feature. We fit an 
elliptical
Gaussian, corresponding to the  synthesised beam, 
to the absorption feature in each plane of the
cube, to obtain an estimate of the  minimum intensity in that
velocity bin. We then used these 
values to construct the
absorption profile in Fig.~\ref{absprof}. 
The position of 
the absorption feature, 
10h 27m 51.3s   -43$^{\circ}$ 54$^{\prime}$ 16$^{\prime \prime}$ 
(J2000), remains constant throughout the velocity range.

The maximum absolute amplitude in absorption 
in one plane (at 2834 \kms ) is
38 mJy, which is only 6\% of the continuum flux 
density of 0.621 Jy~\citep{sk93}
at 20 cm.  
Thus either the gas in the disk is optically thin to HI absorption or
else it has a small filling factor. 

Our beam at 21 cm includes
both the nuclei seen by \nocite{nf95} Norris \&
Forbes (1995)
in 3 \& 6 cm continuum. The flux of the extended emission 
observed at 6 cm is about 280 mJy
which, together with the 621 mJy  observed at 20 cm, 
implies a spectral index of -0.7, and suggests that 
there is no extended 21 cm flux which is unobserved at  6 cm.
The two nuclei have fluxes of 34 and 31 mJy at 6 cm which
implies a flux of about 77 mJy each 
in  20 cm continuum emission, 
and so the HI absorption could result from a 
relatively high (-0.3 -- 0.5) optical depth in front of 
either or both of these nuclei, or could result from 
extended HI absorption covering all the extended 
continuum emission with a low optical depth.
Our data has insufficient resolution to distinguish 
between these alternatives.

Various methods of displaying (e.g. Fig.~\ref{HIrc} ) the cube
indicate that absorption exists over the entire extent of the H~I
velocity range.  Examination of channels before and after the 33
planes of the processed cube suggest that emission and absorption
features exist outside the 555 \kms \ velocity range we
emphasised in our study.

\subsubsection {Mass Estimates}
\label{massest}

 For optically thin line emission 
from a single cloud,
the column density is
\[ \rm N_{HI} = 1.823 \times 10^{18} cm^{-2} 
                    \int T_{spin} \tau (v)\ dv \]
where the integral is equivalent to
\[\rm  \int T_{brightness}(v) \ dv \]
  Integrating the 
flux over the
solid angle of the  galaxy,
the amount of neutral hydrogen mass 
producing the H~I emission is 
 \[ {\mh}_{\rm em} = \left[\ {\rm 23.6 \  \rm v_{systemic}^2 
                       \ h^{-2} 
                       \int_{0}^{\infty} S(v)\ dv} \ \right]\ \mo \]
 
The observed flux density integrated over the velocity range 
(19  Jy \kms; see \S~\ref{emis})
 corresponds to an H~I mass of 
 $\rm 3.5  \times 10^9 \ h^{-2} \  \mo$ (with 75\% in the
tails; \S~\ref{emis}).

For absorbing gas, however, observations are
a combination of T$\rm_{brightness}$ and the temperature
of the continuum source.  Therefore to calculate the mass of this 
gas, we need to assume a spin temperature
and a value of the optical depth ($\tau$)
in the column density equation. We estimate $\tau$
by
following the argument above in Section 3.1.3 and
assuming that the 21-cm continuum follows the same 
spatial distribution as the 6-cm continuum
image of Norris \& Forbes (1995), but scaled to give a total 21-cm 
continuum flux of 621 mJy.

Using the resulting minimum and maximum surface brightnesses (B), we then
determined $\tau$ for each plane in
the cube using
 \[\rm \tau (v) = -ln \ \left[1 \ -\ \ 
\frac{B_{absorption}}{B_{spin} - B_{continuum}}\right]\]
and integrating over velocity in order to
calculate $\rm N_{HI}$ in units of 
atoms cm$^{-2}$. To determine the mass detected by
the ATCA we multiplied by the synthesized beam area and 
converted to solar masses. 
These two approaches 
 to estimating the surface brightness
give a mass range for the absorbing gas
of ${\mh}_{\rm abs} \rm  \simeq [0.2 - 1.0] \  
\times 10^9\ h^{-2} \mo$ for an adopted
 spin temperature of 100$^{\circ}$K.

\subsection {Observations of Globular Cluster Progenitor Candidates}
\label{gcres}
\subsubsection {Spatially Distinct Fragments}

We wish to assess whether there exist H~I fragments in the cube
possessing masses appropriate for GC progenitors.  However we
have not attempted to identify clumps embedded within the tails
as distinct fragments.  Nor have we attempted to distinguish
foreground fragments from tail material with similar velocities.
Therefore our sample of fragments is incomplete and preliminary.

Since the observations emphasized the extended scales ($>$ 30
arcsec) of diffuse emission, we assume that the features which
appear spatially distinct in Fig.~\ref{HIcntr} at the 3 $\sigma$
level (where $\sigma$ =1.3 mJy/beam) are candidate fragments.
(At the 2 $\sigma$ level some features (like A4 and A5) appear
attached to each other and to the tails, however they are clearly
clumpy.)  Some features could, in principle, be artifacts of the
processing procedure. For example, A6, A10, A11 and A12 do not
exist in Fig.~\ref{HIcntr}b, before application of the primary
beam correction. Therefore we assume they could be noise enhanced
in Fig.~\ref{HIcntr}a by the correction and do not include them
in Table~\ref{fragspatial} which lists the sample of fragments
that could be GC progenitor {\em candidates}.

In order to estimate the mass of these candidates we use the
channel maps presented in Fig.~\ref{mosaic} (described in
\S~\ref{velbehav}) along with ATCA spectra. To generate a spectrum of
each fragment, the velocity
plane in which the feature was most apparent was displayed. A 
polygon enclosing the feature was created and the flux density
within this area was then measured in each plane of the cube.  To
be considered as a candidate H~I fragment, a peak
in such a spectrum needed to
\begin{itemize}
\item also exist in the associated panel of the mosaic in
  Fig.~\ref{mosaic}.  In all but one case these features existed
  in a few consecutive velocity planes.
\item have a morphology consistent with the shape of an isolated
  fragment in each velocity plane of Fig.~\ref{mosaic}.  For
  example, it could not be associated with extended or diffuse
  features in any of the consecutive velocity planes.
\end{itemize}  
If features listed in Table~\ref{fragspatial} are noise or are
caused by the processing and analysis procedures, we would expect
to see negative features with the same characteristics as the
apparent emission objects.  However we did not find any negative
features that appeared in consecutive velocity planes in the
profiles.
 
The integrated flux densities were measured from the spectra.
For each fragment this involved subtracting a background emission
value per plane from the value of the fragment; this tabulated
S(v) differs from S(v)-plus-background values by $\sim$10\%.  The
integration included planes that bracketed the detections in
velocity.  For example, if the emission is mainly contained in
one plane then we measured the flux density in that plane plus
the planes on either side of the emission channel (and listed the
number of planes as 3 in Table~\ref{fragspatial}).

The integrated flux density and the velocity associated with the
peak emission of the fragment candidate were used to calculate
the H~I mass via the ${\mh}_{\rm em}$ formula given in
\S~\ref{massest}.

The FWHM velocity range was estimated by fitting a single
Gaussian profile to the spectrum of each fragment (although A5
and A7 were not single peaked in velocity space). Since the main
contribution to the uncertainty, listed in
Table~\ref{fragspatial}, is associated with our selection of the
velocity planes in which the fragment exists, this rough approach
was deemed adequate.  In order to characterise the error in the
FWHM velocity range, we examined the proportionality between the
area under the Gaussian profile and the H~I mass since the
integrated intensity determined from the Gaussian fit should be
consistent with the integrated flux density used to determine the
H~I mass of a fragment.  The difference is less than 20\% for the
single peak fragments.
 
If we assume that the single peaked potential
 fragments are in virial equilibrium, and adopting
a maximum radius, we can estimate an upper limit on their mass.  
The
equilibrium mass is 
\[ \rm M_{virial} = \frac{(velocity \ dispersion)^2 \ 
                 \times \ (mean \ radius)}{G} \]
We use   (FWHM velocity / 2.354)  for the velocity dispersion.  
Since the
fragments are
spatially unresolved, we use the mean radius (11.3 arcsec) of
the synthesised beam as an upper limit on the mean radius  for
each potential fragment.  
Dividing the equilibrium mass  estimate by the H~I mass value 
gives an upper limit on the factor by which the total 
mass of the potential fragment, if it is in equilibrium,
 exceeds the amount of neutral hydrogen.   Along with the
FWHM, flux, and mass values, this ratio is 
listed in Table ~\ref{fragspatial}.   
 
Our detection threshold is 
$\rm  a\ few\ \times 10^6 \ h^{-2}\  \mo$ per plane.  Since the
fragments are measured over at least a few planes, the
lowest mass object we are capable of  measuring in this way is 
comparable to the 
lower mass limit
associated with a globular
cluster progenitor ($\rm \sim  10^{7 \pm 1} \ h^{-2}\  \mo$).    
 
(A fragment appears in the last plane of the contour mosaic,
Fig.~\ref{mosaic}, but not in the zeroth moment map, 
Fig.~\ref{HIcntr}.
If it is not a spurious noise feature,  the  mass 
estimated using the
technique above is 
$\rm  [3.9 \pm 0.8] \times 10^7 \ h^{-2}\  \mo$.)

\begin{table}
\dummytable\label{fragspatial}
\end{table}

\subsubsection{Fragments with Non-circular Velocities}
Fig.~\ref{HIrc} shows regions 
in H~I that have a large
velocity spread  which 
we can  associate with velocity 
anomalies in the optical
velocity field presented in Paper~I. 
However the distinct  \ha
velocity features are usually near the disk of the 
galaxy and, due to the lower resolution of our
H~I data, 
not readily identifiable with isolated H~I intensity features.
We plan to investigate a variety of potential causes for
the velocity anomalies using
visualisation techniques once our data
set has been merged with 
higher resolution baseline configurations.

\section{A Tidal Tale}
\label{tail}
In this section we attempt  to broadly sketch a possible
 history for NGC~3256 by
comparing  its observational 
characteristics with the  
interaction scenarios implied by  
generic hydrodynamical numerical models. 
Since the details of these    models 
do not duplicate the details 
of our observations,  this section  highlights
the need for a computational model designed to mimic this 
particular encounter.

\subsection{Morphological and Star Formation Evolution
During Galaxy-Galaxy Mergers: A Small Selection of Models}

A lesson learned from the classic restricted 3-body study of
dissipationless systems by \nocite{tt72} Toomre \& Toomre (1972) was
that two parent galaxies of equal mass will produce equivalent
counter-arms which grow into tails as the interaction progresses.
Importantly, Toomre \& Toomre also proposed that interactions would
lead to the formation of elliptical galaxies via orbital decay. They
also linked the observed prolific star formation in peculiar galaxy
systems with galaxy-galaxy interactions, suggesting the violent
mechanical agitation would bring a sudden supply of fresh fuel deep
into the galaxy.  NGC~3256, which is observed to have two similar
stellar tails, a common envelope, and an enhanced star formation rate, could
be a ``snapshot'' of two roughly equal-mass disk systems in the
process of merging. Thus ~\nocite{toom77}Toomre~(1977) includes it in
his observation-based schematic of a sequence of merger stages that
begins with NGC~4038/9~(The Antennae).

The NGC~4038/9 close-encounter stage is preceded by an era of
galaxy-galaxy interactions in which significant tidal friction
causes the galaxies to orbit one another with a decreasing
orbital period.  Self-consistent modeling of the galaxies' purely stellar
component \citep{bar92} demonstrates that the decay mechanism
could be independent of any role gas may play. That is, tidal forces
exerted by a companion galaxy on the other galaxy's halo cause
the latter galaxy's orbital angular momentum (from its halo and
subsequently its bulge) to be transferred into spin angular
momentum in its own halo. Through repetition of this loss of
orbital angular momentum, the galaxies' orbits decay and, after
the NGC~4038/9 stage, they fall back together in a rapid series
of closer and closer pericentre passages until the parent galaxy
cores coalesce. This can eventually produce a merger remnant with
an r$^{1/4}$ law profile.

Self-consistent hydrodynamical simulations by
\nocite{mh94}\nocite{mh96} Mihos \& Hernquist~(e.g. 1994, 1996),
which include processes in the ISM, show the star formation
behaviour as two similar disk galaxies merge.  These numerical
experiments illuminate both GC formation and nuclear activity.
\nocite{mh96}Mihos \& Hernquist~(e.g. 1996) find that the inner
structure of galaxies, more than the orbital geometry, determines
the formation and evolution of starbursts in merging galaxies
since dense bulges act to stabilise the galaxies against gas
inflow.  Rather than forming bars, these disks acquire two-armed
spiral features.  The diffusely distributed gas later collects
into a central cloud as the cores begin to merge. The resultant
starburst is brief enough to be comparable to the age of observed
starbursts, and intense enough to be descriptive of
ultra-luminous galaxies.  Additional sites of gas concentration,
and hence possibly sites of enhanced star formation, are the condensations
residing in the well-defined gaseous tidal tails.
 
\subsection{Comparison of   Observations
of NGC~3256 with Numerical Models}
 
\subsubsection{The Initial Encounter}
\nocite{lipari00}Lipari~et~al.\/~(2000) interpret optical observations
as suggesting NGC 3256 is the merger of 3 galaxies.  However the
extended H~I is distributed into 2 broad tails and, although each of
these has substructure, this is consistent with the scenario that
NGC~3256 is mainly the product of two similarly massive parent galaxies
involved in a prograde encounter.  (The fact that the east tail has
60\% less gas than the west tail could indicate that one of the parent
galaxies was less gas-rich since its stellar mass could be similar in
magnitude to the companion's stellar mass.) The extrapolated IR
luminosity of NGC~3256 also supports this scenario.  It is only a
factor of 3 less than that of the ultra-luminous category ($ \geq
10^{12} \cal L_{\odot}$) and it surpasses the 10 $\mu$m luminosity of
typical starbursts by an order of magnitude \citep{grah84}. Hence
NGC~3256 is probably related to ultra-luminous starbursts.
Ultra-luminous IRAS sources tend to have double tails and many have
double nuclei \citep{san88}. If collisions of molecular gas can trigger
events leading to FIR luminosities, then the optimal scenario for
producing ultra-luminous starbursts involves mergers of similarly
massive gas-rich galaxies.  As well, the two radio continuum sources,
which we identify with the original cores of NGC~3256's parent
galaxies in \S~\ref{intro3256}, are approximately equivalent in
luminosity \citep{nf95} which may suggest that galaxies of similar mass
were involved in the interaction.

The substantial spatial width of the H~I tails of NGC~3256 suggests
that our observational viewpoint is almost directly above the orbital
plane of the interaction.  The different shapes of the tails would
then be due to the inclination of the parent galaxies' disks with
respect to the orbital plane, rather than due to extreme mass
differences.  If a galaxy's spin is retrograde with respect to the
encounter orbit, the formation of an extended tail is dampened.  Hence
the existence of two strong tails also implies that the spin of each
galaxy was prograde with respect to the binary orbit.
 
\subsubsection{The Current Encounter Stage}

It is possible that we are currently observing an epoch in which the
parent galaxy cores have already merged and one of our designated
nuclei may simply be a star formation enhancement.  Then this would be
a rather advanced merger stage and the kinematic coupling would
suggest that the remnant system will retain a merger signature
(kinematic and/or structural~\citep{bar92}) that reflects the
characteristics of the close-encounter orbital geometry.  However a
number of observational and theoretical considerations, outlined
below, indicate that NGC 3256 is experiencing the starburst which just
precedes the final core collision when 2 gas rich galaxies with bulges
merge.

\begin{itemize}
\item The common envelope shared by the nuclei in NGC~3256 indicates
that this system has evolved beyond the close-encounter
separation which designates the beginning of an on-going merger.
Also enough time has passed to allow clumps to form in tidal
tails, for enough gas to collect in the centre of each parent to
induce starbursting, and for the galaxy nuclei to approach
closely.
 
\item The {\em global} sense of rotation of the ionised \ha gas and the
neutral hydrogen gas implies that the gas in the inner region of
the system ($\rm H_{\alpha}$) is kinematically coupled to the gas
in the outer region (H~I); compare the velocity field in Paper I
and Fig.~\ref{HIvfld} presented here.  Therefore, although post-pericentre,
the merger has not yet
progressed to a stage where the inner region has lost a large
fraction of its memory concerning its original angular and
orbital momentum.  However, the separation of 5 arcsec
\citep{nf95}, or only about 700 \ho pc, of the northern nucleus
from the southern nucleus suggests that these are parent
galaxy cores which are about to collide.  It seems unlikely that the
observed proximity of the cores is a projection effect since 
the inclination of the galaxy estimated by \nocite{fr78} Feast
\& Robertson~(1978) is only $\rm 40^{\circ}$. 

\item Norris \& Forbes have shown that the supernova rate in each of the two
components is about 0.3 per year, a rate which appears to be far too
high for an extra-nuclear starburst region.  (Since the eastern star forming
feature at 10h 27m 51.7 -43$^\circ$ 54$^\prime$ 13.9$^{\prime\prime}$
(J2000) does not have as strong radio continuum emission we do not
assume it is another nucleus, as proposed by \nocite{lipari00}
Lipari~et~al.\/~(2000).)

\item \citet{lira02} compare the spectral energy distributions of
both nuclei with SEDs of starburst galaxies, QSOs, and low-luminosity
AGN.  The northern nucleus, the brightest X-ray source, is
the heart of the nuclear starburst. However the southern nucleus
has a flat SED between 1-10 microns, indicating that a low-luminosity 
AGN cannot be ruled out. 

\item The H~I absorption feature, spatially coincident with these nuclei,
displays the full range of observed velocities in the system.  There
are a few scenarios which could cause this.  For example, the velocity
range could be evidence of a warp in the H~I disk and that we are
viewing the galaxies' nuclei in projection. But again the morphology
of the tails constrains the warp to relatively low angles of
inclination.  Secondly, if we are indeed observing the orbital plane
almost face-on, the velocity range could indicate that the absorbed
H~I gas resides in extended, inclined tidal loops and fragments
(e.g. \nocite{hm95}Hibbard \& Mihos~(1995)).  However if gas were
being funneled along these features into the central region of
NGC~3256 all the velocities would be redshifted. Thirdly, and this is
the more likely picture, the gas could be streaming. This is either
because the 2 nuclei behave as a dipole or because there exists a
structure such as a bar or an accretion ring (e.g.  
\nocite{kdm93}Koribalski~et~al.\/~(1993)).

\item Star formation activity is strong in NGC~3256, but its gas is not
restricted to the central region. We observe an H~II region
formation rate of 60 times that observed for Sc galaxies occurring
in a region that extends at least 70 arcsec range or about 10 \ho
kpc (Paper~I).  Infra-red data mimics this distribution over
about 4 \ho kpc; most of the IR luminosity arises from outside
the central kpc~\citep{grah87}.  Both the CO \citep{abbj} and
H~I emission are extended with most of the atomic hydrogen gas
residing outside the disk (\S~\ref{emis}).  (The H~I mass in the
centre, if it can be characterized by the gas in absorption, has
an upper limit of about 30\% of the spatially extended H~I gas.)
The onset of the current burst is relatively recent; using 2.5-40
$\mu$m ISO SWS spectra, its age is estimated to be 10 to 20 Myr
\citep{riglutz96}. 

\end{itemize}

These observations are qualitatively consistent with the
\nocite{mh94}Mihos \& Hernquist~(1996) model in which the parent
galaxies have bulges and the onset of enhanced star formation is
delayed until it just precedes the final core collision. As this burst
is occurring, the gas may still appear diffusely distributed and a
small separation is also expected to exist between parent galaxy
nuclei.  We therefore adopt the picture that the designated nuclei are
the cores of the progenitor galaxies and that the nuclei have yet to
collide.

\subsubsection{Timescales}
The proximity of the parent galaxy cores suggest that NGC~3256
has had more than one
pericentre approach. However we estimate the time that has
elapsed by assuming the gas, or ``particles'', at the tip of the
tails were torn from each parent galaxy during the pericentre
associated with the current close-encounter stage.  To estimate
roughly the age of the tails, we assume the sum of the parent
galaxies' masses during the pericentre epoch is the same as that
now observed within the optical envelope of the disk-like part of
the system and the parent galaxies' pericentre separation is
taken to be very small. In Paper I we estimate the I band
envelope mass within 70 arcsec of the \ha nucleus to be $ \rm (2.5
\pm 0.5) \times 10^{10} \ h^{-1} \mo$.  From this mass, and the
radius of the outer tip of the west tail (42 \ho \ kpc), the
characteristic dynamical timescale is about 500 Myr.

To compare models to our data, we need to convert model unit
quantities to physical values.  For the total mass of one parent
galaxy we adopt half of the I band envelope dynamical mass
including its uncertainty since this provides an estimate of the
H~I mass that has been redistributed to the tail during the
interaction ( i.e. $\rm M_{parent} = 0.5 \times 3 \times 10^{10}\ 
h^{-1} \mo$).  The radius which contains half the mass of the
parent galaxy can be estimated from our data using,
\[ \rm r_{1/2} = \frac{1}{2} \ \frac{G \  M_{parent}}{v_{circ}^2} \]
Assuming  that before the interaction the parent 
galaxy had a rotational velocity 
which is comparable to the rotational velocity for 
the currently merging system, we used our estimate of
107 \kms \ (without any correction for the uncertain inclination).

In the Mihos \& Hernquist~(1994) models, unit mass and unit
length are scaled to a specific galaxy using the parent galaxy's
total disk mass and the disk exponential scale length.  A
relationship between the radius at half-mass and the disk scale
length, r$\rm _s$, for an exponential disk can be determined
numerically from
\[
\rm M(R)  =  2 \pi \int_{0}^{R} \Sigma_0 \ r\  e^{(-r/r_s)} \ dr \]
Setting
$ \rm M(R_{1/2}) = \frac{1}{2} M(\infty) $
gives
\[ \rm R_{1/2} = 1.7 \ r_s \]
Defining G $\equiv$ 1, and using the length and mass calculated above, a
unit time converts to about 10 Myr (h = 0.8).  

In Mihos \& Hernquist~(1994), the pericentre through to coalescence
stage spans 50 model units, i.e. 500 Myr, which is comparable to our
estimate of the time elapsed in NGC~3256.  In their scenario for
parent galaxies with dense bulges, a strongly peaked starburst begins
about 40 Myr before the cores coalesce. Hence we consider this model
and our data in adequate agreement.

\subsubsection{The Merger Remnant} 

As noted in \S~\ref{intro3256}, the K luminosity profile is
consistent with the notion that NGC~3256 will turn into an
elliptical galaxy in the final merger state~\citep{mo94}; this
is independent of whether the parent galaxy cores have already
coalesced or not.  The notion of an elliptical remnant receives
some support from the molecular-line survey by \nocite{cdc92}
Casoli et al.\/~(1992) that shows that the ISM of NGC~3256 is not
simply that of a spiral galaxy which is very actively forming
stars. It requires the 
the mixing of the gaseous
components of the parent galaxies, which would occur if
the original material in the galaxies were being redistributed
into the $\rm r^{1/4}$ law associated with ellipticals.

The $\rm 10 \ \mu m $ luminosity of the \ha nucleus rivals that of
Seyfert galaxies~\citep{grah84}, begging the question of whether an
active nucleus currently contributes to the far-infrared luminosity.
This seems unlikely for the northern core since observations do not
provide substantial evidence of an active nucleus~(IR:
\nocite{mo94}Moorwood \& Oliva~(1994) and
\nocite{riglutz96}Rigopoulou~et~al.\/~(1996;ISO-SWS spectra); radio:
\nocite{nf95}Norris \& Forbes~(1995); X-ray:
\nocite{moran99}Moran~et~al.\/~(1999) and \cite{lira02}). We note that
the current starburst ($\rm \sim 3 \times 10^{11}\
L_{\odot}$~\citep{ssp89}) could deplete a significant amount of gas
before this fuel can be funneled into the single dense concentration
required to efficiently stoke an active galactic
nucleus~\citep{san88}.  Additionally, although NGC~3256's luminosity
may increase as the parent galaxies' nuclei more closely approach each
other, the hydrodynamical simulations indicate that active nuclei are
unnecessary for explaining ultra-luminous FIR
emission~\citep{mh96}. \citet{lira02} examine the AGN scenario in
detail, concluding that the powerful X-ray emission could be driven
solely by the current episode of star formation, which supports the
starburst scenario.  However they cannot rule out the existence of a
{\em low-luminosity} AGN in the southern core.
 
\section{Discussion}
\label{discus}
\subsection{Globular Cluster Formation in Mergers}
\subsubsection{H~I Fragments as Globular Cluster ``Cradles''}

Some likely sites for globular cluster formation during mergers
include detached H~I fragments and the giant H~II regions (see
Paper~I) and gas clumps that form in the tails of merging
galaxies. Examples that dwarf-galaxy-sized
fragments form in tidal features include the two condensations
measured by \nocite{hvg94} Hibbard et al.\/~(1994), in the tails
of NGC~7252, with an H~I mass of about $\rm 10^8 \ h^{-2} \ \mo$.
\nocite{dm94} Duc \& Mirabel (1994) have also found two dwarf galaxies
with masses on the order of 10$^9 \mo$ in the tidal debris of
Arp~105. Optical studies of NGC~3256 also are revealing stellar
substructure in the tails \nocite{Kniercharl01}(Knierman et al.
2001).  However in our study of NGC~3256 we only designated
spatially distinct clumps as GC progenitor candidates.

In this paper we focus on a GC formation scenario related to
the \nocite{sz78} Searle \& Zinn~(1978) idea that globular
cluster formation occurs in fragments around collapsing
protogalaxies.  In the merging of a pair of galaxies, gas
fragments associated with the tails and loops of the merging
system form 
stellar subsystems (eg. \nocite{elm97}Elmegreen~(1997)) and papers
therein).  These fragments, together with the globular clusters
that formed within them, subsequently become part of the final
merged system.  Simultaneously, galaxy interactions detach gas
fragments from the parent galaxies.  Evidence for detached
fragments in merging systems comes from \nocite{hut89} Hutchings'
(1989) study of IRAS galaxies.  Fifty percent of his systems are
interacting and many display weak subcomponents in
their H~I profiles.  Such profile components are unlikely to be
associated with the individual members of the interacting pair,
and he suggests that they are due to material detached from the
galaxies during the interactions.

Fragments which were not immediately captured by the emerging galaxy
may evolve into the dwarf galaxies observed today, such as nucleated
Blue Compact Dwarfs (BCD) and dwarf ellipticals~(e.g. NGC 1705,
\citep{kcf93,meur92}) and dwarf spheroidals~\citep{zin93}.  The dwarf
galaxy associated with the tidal tail of Arp 105~\citep{dm94} provides
direct evidence for this type of scenario. If we assume that the star
formation efficiency is $\sim 10$\%, then the masses of these dwarfs
suggests that the minimum H~I mass for GC formation in fragments is
$\rm 10^{7 \pm 1} \mo$.  Hence the scenario is that fragments which
are torn from interacting galaxies and which satisfy this mass limit
will fall back towards the potential well of the remnant galaxy and
experience tidally induced star formation.  These episodes of
starburst could be triggered by tidally generated gravitational
instabilities~\citep{ekt93}, by interactions between pairs of H~I rich
clouds~\citep{bri90}, or by turbulent compression~\citep{elm97}. The
clusters-to-be can be identified with the nuclei of the BCD dwarf
galaxies~(e.g. \nocite{kcf93} Freeman~(1993)) or the progenitors of
the GC systems observed in dSph's (e.g. the Fornax Dwarf contains 5
GC's)~\citep{zin93}.  The diffuse envelope surrounding the GC
progenitors would be stripped off in the interaction, and the BCD
nucleus survives as a GC when these ``cradles'' are subsequently
accreted by the emerging elliptical.

From our preliminary analysis, NGC~3256 appears to have at least
3 spatially distinct H~I fragments with H~I masses which fulfill
the minimum mass criterion; see Table~\ref{fragspatial}.
Inspection of smoothed Digitized Sky Survey images suggest
Fragments A2 and A7 could have stellar components, but deeper
optical images are needed to confirm this.

One could plan to test the plausibility of various GC
formation scenarios by determining whether the observed numbers
of GC progenitor {\em candidates} at every stage of the
interaction-through-merging sequence are consistent with the
statistical properties of small-scale features (eg. condensations
of particles) in the hydrodynamical models of merging galaxies.
These H~I progenitor candidates in NGC~3256 could be examined for
kinematic consistency with the models' small-scale features.

The estimated virial mass of each fragment is similar to its H~I mass.
Caveats about this estimate include the short timescale in which the
gas must virialize and the possibility that tides contribute to the
observed velocity width.  However if the gas is in equilibrium, then
the fragments do not contain a substantial dark component and could be
torn from a region of the parent galaxy which does not contain much
dark matter, or at least that they are torn only from the gaseous
component.  This is not in contradiction with the observation that
many dwarf galaxies contain a substantial fraction of dark matter (see
\nocite{gw94} Gallagher \& Wyse~(1994)).  Among the local Group dwarfs
(\nocite{mateo98}Mateo~1998) there are several examples of dwarf
galaxies with globular-cluster-like $ M/L$ ratios ($ M/L \sim 2$).

\subsubsection{Enhanced Star Formation
and  Globular Cluster ``Cradles''}
 
We may have expected to observe giant H~II regions within the
merger produced H~I fragments if we also happened to be observing
an epoch of enhanced star formation within the fragment.
\nocite{elm97}Elmegreen \& Efremov (1997 and papers listed
within) propose that stars form efficiently when the velocity
dispersion is high since the binding energy per mass will also be
high.  The dispersion, c, within gas in a turbulent
and virialized cloud is $\rm c \propto (PM^{2})^{1/8}$. Using
typical values for c and mass (M) for giant molecular
clouds (3.8 \kms \ and a few times $\rm 10^5 \mo$;
\nocite{harpud94}Harris \& Pudritz~1994) we can compare the 
pressure, P,  in giant molecular clouds with the P in
NGC~3256's fragments. Using the average
c for NGC~3256's fragments (excluding A5 \& A7)
and a mass of $\rm 10^7 \mo$, indicates a fragment's P could be 
$\rm 10^4$ times higher that that of a molecular cloud.
This is a condition that Elmegreen \& Efremov suggest 
will allow the formation of bound clusters containing 
massive stars. 

A fragment's appearance during the star formation episode which
produces such clusters may be similar to that of an H~II
galaxy~\citep{dm94} or a nucleating dwarf elliptical such as NGC~1705.
However it would be serendipitous to observe an epoch of H~II region
formation in our few detached fragments because the lifetime of a
giant H~II region is only $\sim$5 Myr compared with the duration of
the galaxy interaction ($\sim$500 Myr).  Although we require sensitive
optical observations in order to determine if the HI fragments contain
stars, there is evidence that candidate stellar dwarfs are embedded in
the tidal tails~\citep{Kniercharl01}.  The origin of these tail
substructures may be understood from the suggestion by Elmegreen et
al.\/~(1993) that, during galaxy interactions, clouds form by
gravitational instabilities in the outer part of the disks; the
interaction may also produce dense cores in molecular clouds and
trigger the H~II region formation by external
compression~\citep{elm93}. These clouds and their associated H~II
regions may be ejected from the disks and evolve into dwarf galaxies.
The \citet{ekt93} star formation scenario could also link together the
ideas that \g form in tidal fragments, presented in this paper, and
that \g could form in giant H~II regions, which we examined in Paper
I.

\subsubsection{Specific Frequency and HI progenitors}

The \sn of globular clusters, which is the number of \g per unit
absolute magnitude $M_V = -15$ of galaxy luminosity, is such that
(i.e. non-cD) elliptical galaxies have more than twice as many \g per
unit stellar mass as spirals (\nocite{az93}Ashman and Zepf~1993).
Thus in Paper I we calculate that between 40 and 80 \g need to be
created if NGC~3256's final state is to be identified with an
elliptical remnant. We also show that, throughout the merger
timescale, NGC~3256 will produce a sufficient number of H~II region
progenitors, as in the \nocite{kc88} Kennicutt \& Chu (1988) scenario,
to create the required GC's.  However we also believe that \g with H~I
fragment progenitors, as in the Searle \& Zinn scenario, contribute to
some fraction of the \s.

We are encouraged in this notion by the evidence that the progenitor
of the globular cluster $\omega$~Centauri evolved from a satellite
galaxy. As well as displaying a flattened shape and rotation
(\nocite{vanleeuwen2000}van Leeuwen \etal~2000), $\omega$~Cen's heavy
element abundance ratios, in the bimodal calcium abundance
distribution, favour a scenario in which the ejecta from an earlier
generation of stars enriched the next generation.
\nocite{norfreemig96} Norris \etal~(1996) suggest that in order for
the proto-cluster to have not been disrupted by the Milky Way it had
to evolve away from the central dense regions of the Galaxy.  This
condition can be satisfied if the progenitor was a halo gas fragment
of $\rm 10^8 \mo$ that evolved into a satellite galaxy and was
subsequently disrupted, as in the Searle and Zinn picture.
\nocite{dinescu99}Dinescu \etal~(1999) elaborate on the disruptive
final stage which removes the dwarf galaxy's envelop of stars,
claiming from their kinematic studies that $\omega$~Cen could have
originated in a strongly retrograde satellite as massive as the
Sagittarius dwarf spheroidal.  We apply this scenario to the case of
interacting galaxies by noting that early disruption can also be
reduced if the gas fragment progenitor was originally far-flung tidal
debris.

Since it is beyond the scope of this paper to analyze clumps
embedded in the tidal tails and we lack knowledge about the longevity,
and other characteristics, of our observed detached H~I fragments, we
cannot estimate the number of \g that will form in the NGC~3256 merger
by employing the analysis strategy used on H~II regions in Paper I.
Perhaps a hydrodynamical simulation specifically of the NGC~3256
interaction sequence would be illuminating.  However, using a general
interaction model can aid speculation, as the following example
demonstrates.
 
 In an interaction of 2 equal mass parent galaxies by \nocite{bh96}
Barnes \& Hernquist (1996) numerous condensations occur in the tidal
tails of the parent galaxies.  These simulated clumps include gas
since 10\% percent of each galaxy's disk in the model consists of gas
which is radiatively cooled until the temperature drops below $\rm
10^4$K.  This non-gravitational process means that the model cannot
readily be scaled to NGC~3256 and that the gas approximates the warm
ISM rather than our cool fragments. Thus this model may be more
relevant to the \nocite{ekt93}Elmegreen \etal (1993) scenario,
mentioned at the end of the previous section, than to pure H~I
fragments.  However, as a first approximation, let us assume that tail
condensations are related to our fragments, and that their numbers and
fragment masses do not change with size of the parent galaxies.

To make a generous estimate of the number of condensations that form,
we need to assume that both galaxies are in prograde orbits and that
their disks are not inclined with respect to the orbital plane.  Then
each might generate 2 dozen bound clumps, most forming shortly after
the tails start to develop.  \citet{bh96} state that most of these
form in the outer region of the tails and many have high enough
orbital periods to survive tidal disruption. Visual inspection of the
number of clumps in the outer half of the tidal tails in their Fig.~20
suggests that up to 50\% are still bound after 2 Gyr.

Although the bound clumps in the simulation originate with stellar
overdensities which subsequently accumulate gas, the stellar component
can be tidally dispispersed while the more compact gas component
resists disruption.  Initially the condensations have an approximate
mass range of $\rm 10^{7-8} \mo$, 25\% of which is gas. This suggests
that about 20 condensations, once stripped of their stars, have the
appropriate mass range for our scenario and may survive, thus
potentially providing up to a half of the \g required by specific
frequency arguments in the case that NGC~3256 evolves into an
elliptical remnant.

However to assume that these $\sim$20 gaseous proto-\g are relevant to
NGC~3256's interaction is optimistic.  For example, only if the clump
mass does not scale with parent galaxy size will all of the simulated
gas condensations contain enough mass to be identified with the
observed NGC~3256 fragments.  In the model the number of condensations
in the tail is this high only if the parent disks are not inclined to
the orbital plane and the gas density in each dwarf is high because it
is collected from a thin ridge line in the tails. However NGC~3256's
parent galaxies are likely to have some inclination and we observe
that the gas tails are broad. Further, to be identified with H~I
fragments described in our paper, some of the simulated condensations
should have detached themselves from the stellar tidal features such
as loops and tails within 500 Myr.  So while these general models of
\citet{bh96} support in principle the formation of \g from tidal
debris, clearly more fitting simulations are required. These and
multi-wavelength observations of the characteristics of gas
condensations and fragments are required before we can calculate a
realistic contribution of tidal debris dwarfs to the \s of merger
remnants like NGC~3256.

\section{Conclusions}
Our H~I observations of NGC~3256 
\begin{itemize}
\item Confirm that this system was mainly formed as the result of
a prograde encounter between two similarly-sized gas-rich spirals.
The cores of these parent galaxies appear to be on the 
point of coalescing.  We estimate that the encounter that 
produced the tidal tails occurred 500 Myr ago. 
\item Show $\rm [3.5 \ to \ 4] \times 10^9 \ h^{-2} \  \mo$ of neutral
hydrogen gas is  providing fuel for highly
 IR-luminous starburst activity. However this gas is
distributed over several kpc, despite the conventional assumption that
ultraluminous galaxies are powered by a compact 
circum-nuclear starburst.

Also the H~I in the central region has not yet been depleted by
the activity.  An absorption feature, which displays the same
velocity range as the rotation curve, is consistent with
streaming gas.

\item Display a number of apparently isolated fragments. The H~I
masses of these ($\rm \geq 10^{7 \pm 1} \mo$) suggest they could be
progenitors of globular clusters.  These may pass through a dwarf
galaxy stage during their evolution.  The equilibrium and H~I masses
of these fragments are similar which suggests they were torn from a
portion of the system which does not have dark matter. The absence of
dark matter is also a characteristic of globular clusters.

\end{itemize}

Both lower and higher spatial resolution ATCA observations of NGC~3256 in H~I
spectral-line mode have been acquired from which we examine the
velocity anomalies and morphology of the inner region of this
system. When these data are merged with the set presented here,
the resolved absorption feature will allow us to examine whether
the gas is streaming between nuclei, along a bar, or in an
accretion torus.  These data will also allow us to compensate for
the incompleteness of our H~I fragment sample and assess whether
stellar enhancements in optical data are associated with these
fragments.

\section{Acknowledgments}

We would like to thank 
Jacqueline Van Gorkom facilitating this
collaboration and Judith Irwin and Claude Carignan 
for for helpful instruction.  We are grateful to
 Peter McGregor and Sung-Eun 
Kim for allowing us to publish their image and IR results in
advance of publication. We also wish to thank L. Johansson, 
S. Aalto, and J. Black 
for allowing us to reproduce  data on Fig.~\ref{4bd}.
We also thank Baerbel Koribalski for stimulating discussions and for leading
an enlightening discussion involving several other colleagues.


One of us (J.E.) thanks Baerbel Koribalski and the ATNF for their
support which has contributed to the completion of this paper.
J.E. acknowledges the support of an Australian National University
Postgraduate Scholarship. 
This research has made use of the NASA/IPAC Extragalactic  Database
(NED) which is operated by the Jet Propulsion Laboratory, Caltech, 
under contract with the National Aeronautics and Space 
Administration. The Australia Telescope is funded by the
Commonwealth of Australia for operation as a National Facility managed
by CSIRO. We also thank the STScI and PPARC for use of the SkyView tool.

\newpage

\figcaption[EngNor.fig1]
{An overlay of 3 different data sets of NGC~3256.  Data from the
  neutral hydrogen spectral line cube is presented as an intensity
  zeroth moment map in red.  This data was acquired with the Australia
  Telescope Compact Array (ATCA) and consists of only 2 antenna
  baseline-configurations. The I band (coloured green) and ionised
  hydrogen (coloured blue) images were acquired with the 1 metre
  telescope at Siding Spring Observatories.  Note where the radio and
  optical data overlap the false colours mix to form yellow.  North is
  to the left and east to the bottom.  Fig. 8 indicates the scale
  associated with the ATCA data.
\label{postcard}} 

\figcaption[EngNor.fig2]
{The Southern Source in NGC~3256. The lower panel 
is McGregor \& Kim's~(1994)  K image acquired
with the CASPIR detector on the 2.3m telescope at Siding 
Spring Observatory.  
The upper image consists
of 3 different data sets.  The images have been shifted to the 
coordinates of our \protect\ha CCD image (see Paper I)
and overlayed using IRAF's 
{\bf rgbsun} task.  
McGregor's  K image is designated red, our
continuum-subtracted \protect\ha data is assigned blue, and
the  3cm continuum
emission observed by Norris \& Forbes~(1995) green. Where the data
sets overlap the colours add together. For example,  
the prevalence of
the IR emission makes  the H~II regions appear magenta and   
the 3 data sets coincide on the northern nucleus rendering it white.
Since the southern ``nucleus'' is obscured in \protect\ha it appears
bright yellow.  
The circles mark the peak and the tip, i.e.  edge,  of the extended 
$\rm ^{12}CO \ J=2-1$ line  emission
observed by Aalto et al.~(1991); their radius of 5 arcsec corresponds
to the SEST pointing error. (The CO tip occurs at the position of the
HII region designated B4 in the photometry figure 
presented in Paper I.) We thank
McGregor, and Norris \& Forbes, and Aalto et al. for permitting us 
to reproduce their data in this format. \label{4bd}} 

\figcaption[EngNor.fig3.ps]{The Parkes Profile of 21-cm 
emission in NGC~3256. 
The emission combined with the absorption of the background
 continuum sources produce  
the Parkes flux density values displayed by the solid-line. 
The heliocentric velocities on the bottom axis were 
calculated using the radio definition of velocity while those
on the top are calculated using the optical definition.  The
dashed-line displays the flux density values of the ATCA data
once it has been
multiplied by a Gaussian with a  FWHM of 15 arcmin in order
to mimic the Parkes data (see \S~\protect\ref{emis}). 
\label{pksprof}} 

\figcaption[EngNor.fig4.ps]{HI contour mosaic.
Pairs of planes from  the 33 plane
21-cm spectral line cube have been summed resulting in 
a velocity width  of about 34
$\rm km \ s^{-1}$ per plane. The r.m.s. 
noise in each plane is  1.3 mJy;  and the contour range, 
in $\sigma $ levels, is -60,-50,-30,-15,-8,-6,-5,-4,
-3,3,4,5,6,8,10,12.
The beam size is displayed in the top 
lefthand corner of the first panel.
\label{mosaic}}

\figcaption[EngNor.fig5.ps]{The distribution of H~I in  NGC~3256.
An integrated surface brightness (zeroth moment) map
of the primary beam corrected ATCA H~I cube. 
The data cube was smoothed 
in the spatial direction (on
the scale of the width
of the short axis of the synthesised beam)
and smoothed 
in the velocity direction
(over 5 planes).  The calculations do not use
intensities between -1.5 sigma and 1.5 sigma from the background
since these intensities were taken to be noise.
As another 
conservative measure we 
subsequently {\em blank}ed
features which
did not appear  simultaneously in the zeroth, first, and
second moment maps and 
which were much smaller than
the synthesised beam.  The integration occurs
 over a  velocity range of 555 \kms .
The labelled 
spatially separated fragments
also appear in a 1.5 $\sigma$ moment map from the cube before
the primary beam correction has been applied; for the 
characteristics of the fragments see 
Table~\protect\ref{fragspatial}.
\label{HIcntr}
The figure on the right is the same
except calculated using the
ATCA H~I cube before applying the primary beam correction.
}

\figcaption[EngNor.fig6.ps]
{The ATCA Profile of 21-cm emission in NGC~3256.
The solid-line spectrum does not include measurements of 
the absorption
feature. The error bars reflect the uncertainty in the measurement due
to the r.m.s. and the 
use of different display parameters.  The dashed-line spectrum 
portrays 
ATCA data as Parkes single-dish telescope measurements 
(see $\S$~\protect\ref{emis}).  
The heliocentric velocity has been calculated
using the optical definition.
\label{emprof}}

\figcaption[EngNor.fig7.ps]
{HI velocity curve.  A position-velocity plot of 
the ATCA cube along the same position angle (90$^{\circ}$
and at the same
declination as the
ionised hydrogen velocity curve presented in Paper~I.
\label{HIrc}}

\figcaption[draftjuly98.fig8]{The H~I velocity field in NGC~3256.
Artifacts 
suspected of being noise have been blanked out in this 
mean velocity (first
moment) map of the primary beam corrected
ATCA 21-cm spectral line cube. Values between -.0013 and .0013 Jy
(i.e. $\sim 1.5 \sigma$) 
were not used in the moment analysis. The 
mean diameter of the synthesised beam is 23 arcsec.
 The 
full colour range was chosen to maintain the convention 
`red corresponds to redshift and
blue corresponds to blueshift'.  To create the intervening velocity range,
the red and blue extremum were blended together.
(Note: The saturation of the 
red was set such that it was dull and the blue 
chosen was a `warm', bright blue.  Hence the red colour, in terms
of human visual perception, 
appears to recede and the 
blue colour visually jumps forward, 
supporting the kinematic meaning of 
redshift and blueshift.)   
\label{HIvfld}}

\figcaption[EngNor.fig9.ps]{The ATCA Profile of the Absorption
 Feature in NGC~3256.
The flux densities were determined using by fitting a Gaussian to
the unresolved absorption feature in each plane of the cube.  The 
uncertainty is dominated by the error in this fit.
\label{absprof}}

\end{document}


\tablenum{2}
\begin{deluxetable}{ccccccccc}
\tablecaption{H~I spatial fragments around NGC~3256.\label{fragspatial}}
\tablecomments{The fragments are
labeled on the H~I contour plot. The J2000 positions coincide with the 
peak H~I intensity and have an error
of 8 arcsec (or 2 pixels). The heliocentric radial velocity is also
associated with the plane containing 
the peak of the emission. The flux density is the mean
flux density over the number of planes in which the fragment exists. 
The velocity spread, FWHM,
is parameterised by the FWHM of a Gaussian fit to each HI spectrum
and has an estimated error of about 20\%. The virial mass is
an upper limit which assumes a spatial mean radius of 11.3 arcsec
for every fragment;  the other  formal
error contribution is the error in the velocity dispersion.  The 
factor by which the virial mass exceeds the H~I mass is also
an upper limit.  
Fragments A5 and  A7 are multi-peaked in velocity space.}
\scriptsize
\tablewidth{0pc}
\tablehead{
\colhead{} &
\colhead{A1} &
\colhead{A2} &
\colhead{A3} &
\colhead{A4} &
\colhead{A5} &
\colhead{A7} &
\colhead{A8} &
\colhead{A9} \nl
}
\startdata
 Position & & & & & & & & \nl
 $\alpha$ (hrs) & 10 27 58.71 &10 27 54.98 &10 27 52.40 &10 27 39.08 
                &10 27 37.60 &10 27 35.74 &10 27 59.79 & 10 28 06.45 \nl
 $\delta$ (deg) & -43 58 44.0 & -43 49 24.0 &-43 56 32.0 & -43 50 32.0 & 
                 -43 51 16.0 & -43 55 12.0 & -43 51 36.0 & -43 51 15.9 \nl
&  & & & & & & & \nl
Radial Velocity (\kms) & 2666.1 & 2884.4 & 2867.6 & 2733.2 & 2733.2 & 
                       2750.0 & 2699.6 & 2682.9 \nl
No. of Planes &   10  &   6  &   6  &   7  &   7  &  10  &   7  &   3 \nl 
FWHM (\kms) &  53.  & 29.  & 35.  & 59.  & 70.  & 145. & 43.  & 24.  \nl  
&  & & & & & & & \nl
Flux Density (Jy) &  0.0052  & 0.0035 & 0.0030  & 0.0024 &
 0.0033 & 0.0043 & 0.0031 & 0.0060 \nl
$\pm$ (Jy) & 0.0004 &  0.0011 & 0.0002&  0.0002 &  0.0004 &  0.0003&   0.0004&   0.0016 \nl
&  & & & & & & & \nl
H~I Mass  &  & & & & & & & \nl
$\rm \times 10^7 \ h^{-2} \ \mo $ &  14.6 $\pm $  .2 &  7.0  $\pm $ 2.3 &  5.8  $\pm $ 0.4 &  4.9  $\pm $ 0.5 &
  6.9  $\pm $ 0.9 &  12.8 $\pm $ 0.8 &  6.2  $\pm $ 0.7 &  5.1  $\pm $ 1.4 \nl
&  & & & & & & & \nl
Virial Mass &   & & & & & & & \nl
$\rm \times 10^7 \ h^{-1} \ \mo $ &  $<$ 17.0 &  $<$ 5.5  &  $<$ 8.2  &  
                                   $<$ 21.7 &    -       &      -     &  $<$ 11.3 &  $<$ 3.7  \nl
&  & & & & & & & \nl
$\rm {\left (\frac{Virial \ Mass}{H~I \ Mass}\right)}_{max}$ &  & & & & & & & \nl
$ \rm h^{-1}$ & 1 &  1 &  1 &  4 & - &  -  &    2 &  1 \nl
\enddata
\end{deluxetable}